\documentclass[9pt,twocolumn,twoside]{optica}
\setboolean{shortarticle}{true}
\setboolean{minireview}{false}
\pdfoutput=1

\title{Quasi Bound States in the Continuum with Few Unit Cells of Photonic Crystal Slab}

\author[1]{Alireza Taghizadeh}
\author[1,*]{Il-Sug Chung}

\affil[1]{Department of Photonics Engineering (DTU Fotonik), Technical University of Denmark, Building 343, DK-2800 Kgs. Lyngby, Denmark}

\affil[*]{Corresponding author: ilch@fotonik.dtu.dk}

\dates{Compiled \today}

\ociscodes{(230.5298) Photonic crystals, (140.3945) Microcavities, (260.2030) Dispersion, (140.0140) Lasers and laser optics.}

\doi{\url{http://dx.doi.org/10.1364/optica.XX.XXXXXX}}

\usepackage{graphicx}
\graphicspath{{./images/}}
\usepackage{commath, amsmath, amssymb}

\usepackage{fancyref}
\usepackage{xcolor}
\usepackage[normalem]{ulem}
\newcommand{\new}[1]{\textcolor{blue}{#1}}

\begin{abstract}
Bound states in the continuum (BICs) in photonic crystal slabs represent the resonances with an infinite quality(Q)-factor, occurring above the light line for an infinitely periodic structure. We show that a set of BICs can turn into quasi-BICs with a very high Q-factor even for two or three unit cell structures. They are explained by a viewpoint of BICs originating from the tight binding of individual resonances of each unit cell as in semiconductors. Combined with a reciprocal-space matching technique, the microcavities based on quasi-BICs can achieve a Q-factor as high as defect-based PhC microcavities. These results may enable experimental studies of BICs in a compact platform as well as realizing a new concept of high-Q mirrorless microcavities.
\end{abstract}

\setboolean{displaycopyright}{true}

\begin{document}
	
\maketitle
\thispagestyle{fancy}
\ifthenelse{\boolean{shortarticle}}{\abscontent}{}

Bound states in the continuum (BICs) are peculiar solutions of wave equations, which are spatially bound and spectrally discrete with an infinite lifetime, equivalently an infinite quality factor (Q factor), at frequencies of continuum of unbound modes. The concept of BIC was first proposed as a solution to Schr{\"o}dinger equation with a complex artificial potential \cite{Neumann1929}. Since then, many different types of BICs have been reported in various physical systems including quantum \cite{Bulgakov2011}, acoustic \cite{Groves1998, Linton2007}, water \cite{Linton2007}, and photonic systems \cite{Marinica2008, Weimann2013}. Recently, BICs in photonic crystal (PhC) slabs have attracted substantial attention as a platform for studying interesting phenomena, e.g. topological charges, as well as a new way of confining light in surface-normal direction ($z$-direction), allowing for novel designs of photonic devices \cite{Lee2012, Hsu2013, Yang2014, Zhen2014, Blanchard2016}. The robustness of BICs in PhC slabs over structure variations as a result of the conservation of topological charges is an important advantage in experimentally implementing BICs \cite{Zhen2014}.  

BICs in PhC slabs occur above the light line, i.e., in the leaky regime of a resonance frequency $\omega$ versus in-plane wavevector $\mathbf{k}$ relation (frequency dispersion), when specific conditions are met. The conditions are that the leaky modes in the slab have a mismatching symmetry from the free-space modes of surrounding materials (symmetry-protected BICs), or multiple leaky modes destructively interfere with each other in the far-field (non-symmetry-protected BICs) \cite{Yang2014}. These conditions are ideally valid for an infinitely periodic structure at specific in-plane wavevector $\mathbf{k}_\text{BIC}$, which means the Q-factor in the surface-normal direction $Q_\perp$ is high only in the vicinity of $\mathbf{k}_\text{BIC}$.
The Fourier-transform of a BIC field profile ($k$-space mode profile) which is extend over an infinite PhC slab, is a delta function at $\mathbf{k}_\text{BIC}$.

\begin{figure}[t!]
	\centering
	\includegraphics[width=0.4\textwidth]{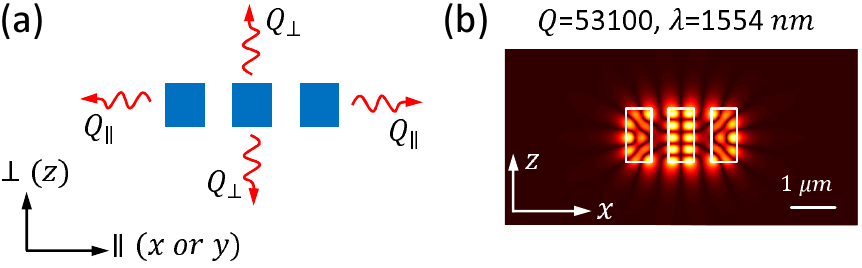}
	\caption{\new{(}a) A schematic cross-section of a microcavity formed by a few unit cells of a PhC slab. (b) The normalized fundamental mode profile ($|E_y|$) of a 2D microcavity with 3 unit cells.}
	\label{fig:CavitySchematic}
\end{figure}

A motivating question of this work is what happens to the BIC as the spatial extension of a PhC slab reduces to a few unit cell structure. In such a finite structure as illustrated in Fig. \ref{fig:CavitySchematic}(a), the $Q_\perp$ is no longer infinite due to a finite-width $k$-space mode profile. The in-plane loss via terminations can be enormous due to a short propagation time over a few unit cells, leading to a very small in-plane Q-factor $Q_\parallel$, which makes the entire Q-factor considerably small since $1/Q=1/Q_\perp+1/Q_\parallel$.  In this Letter, we show that a set of BICs can turn into quasi-BICs with a fairly high Q-factor even for two or three unit cell structures, as shown in Fig. \ref{fig:CavitySchematic}(b). These BICs feature a strong resonance in individual unit cells, which results in an ultraflat frequency dispersion and a very slow in-plane group velocity, abruptly reducing the in-plane loss. The origin of this phenomenon is different from that of the slow light in conventional PhC slab modes below the light line, and is discussed in analogy with the tight-binding of individual atoms in semiconductors. Furthermore, we propose a method to match the $k$-space mode profile of a quasi-BIC with the Q-factor dispersion (Q factor versus in-plane wavevector) of a BIC defined for an infinite PhC slab, which reduces the surface-normal loss by a few orders of magnitude. 
These results may provide an insight for designing compact BIC platforms for experimental studies as well as enabling new high-Q  microcavities for various applications, which operates above the light line without a surrounding mirror. These quasi-BIC based microcavities could be an alternative to the defect-based PhC microcavities, which work below the light line, i.e., relying on the total internal reflection.

\begin{figure}[t!]
	\centering
	\includegraphics[width=0.4\textwidth]{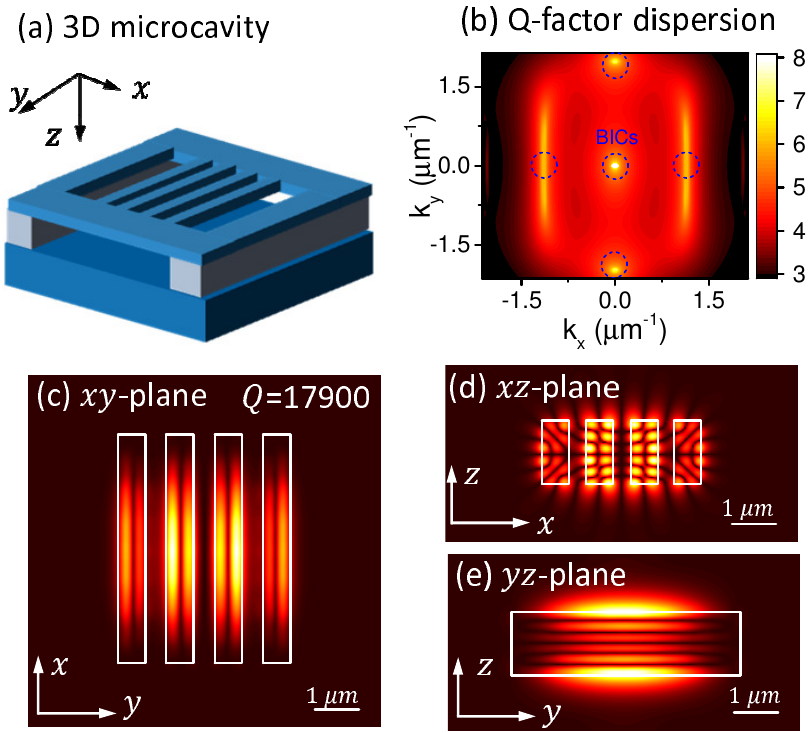}
	\caption{(a) Schematic of a 3D microcavity with 4 unit cells. (b) Q-factor dispersion contour [$\log_{10}(Q)$] of the PhC design for Fig. \ref{fig:Field3DCase}(a). (c)-(e) Normalized field profiles ($|E_y|$) of the fundamental TE-like mode for the 3D cavity of Fig. \ref{fig:Field3DCase}(a). 
	\label{fig:Field3DCase}}
\end{figure}

In this study, 1D PhC slabs of bars, also known as \textit{high-contrast grating} \cite{Chang2012, Taghizadeh2015, Park2015} are used for more transparent understanding. However, we expect that the all proposed concepts are applicable to the 2D PhC slabs of rods or holes. The structure dimensions are scaled for a resonance at the communication wavelength of 1550 nm. It is assumed that the PhC slab is made of Si with a refractive index $n$=3.48 and suspended in air, and the Si bars are infinitely long (2D simulation, and $\mathbf{k}_\text{BIC}$=$k_\text{BIC}\hat{x}$). 3D structures are also considered in Figs. \ref{fig:Field3DCase}(a) to \ref{fig:Field3DCase}(e) and S4(a) to S4(d).  Details of numerical technique are provided in the Supplemental 1.

\begin{figure}[t!]
	\centering
	\includegraphics[width=0.45\textwidth]{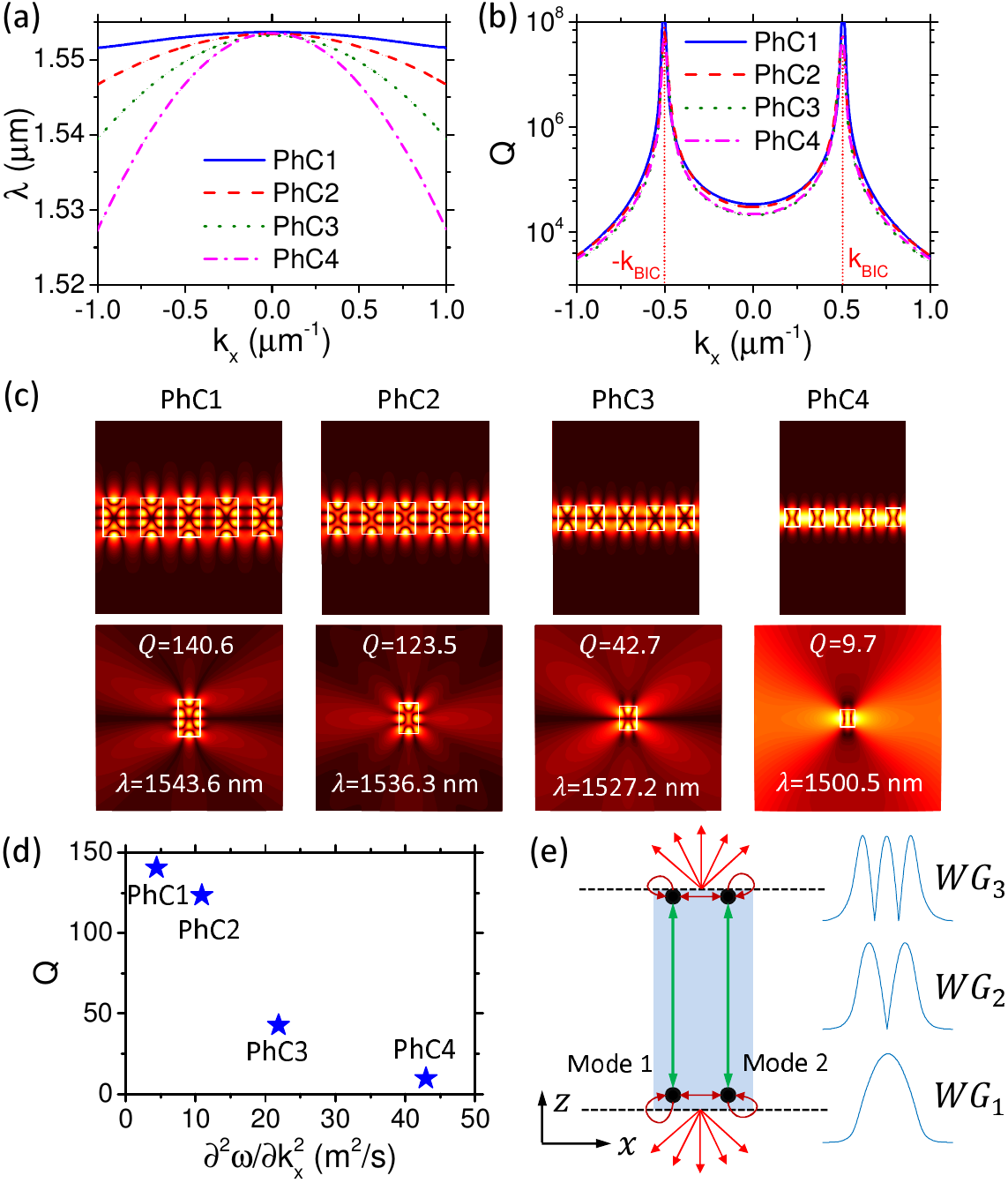}
	\caption{(a) The wavelength dispersion and (b) Q-factor dispersions for PhC1 to PhC4, all with infinite periodicity. (c) The normalized mode profiles ($|E_y|$) of infinite PhC structures (top row) and the corresponding resonances of a single unit cell (bottom row) (d) The Q-factor of the resonance of a single unit cell versus the frequency dispersion curvature $\partial^2\omega/\partial k_x^2$ of an infinite number of unit cells. (e) Schematic of a single unit cell, seen as a finite length waveguide in the $z$-direction with several guided modes WG$_i$. Mode 1 and 2 can be any of WG$_i$ with the same spatial symmetry, e.g., WG$_1$ and WG$_3$ in this example, and they interact with each other and continuum of modes in the surrounding region at the two interfaces. 
		\label{fig:MieResonance}}
\end{figure}

Firstly, let us discuss the relation between the ultraflat dispersion of an infinite PhC slab and the strong individual resonances in each unit cell. The single unit cell Q-factor of a subwavelength structure typically ranges from 1 to 10 \cite{Landreman2016, Campione2016}. In contrast, the single unit cell resonance associated with the ultraflat-dispersion BICs exhibits one or two orders of magnitude larger Q-factor of which origin will be discussed below. 
To illustrate the correlation between these BICs and the strong resonance of a single unit cell, four different PhC structures (PhC1 to PhC4) are compared. 
As shown in Figs. \ref{fig:MieResonance}(a) and \ref{fig:MieResonance}(b), the PhC structures are designed to have almost identical bandedge wavelengths at 1553 nm and non-symmetry-protected BICs peaked at $k_\text{BIC}\approx0.5~\mu\text{m}^{-1}$, while having different dispersion curvatures $\partial^2 \omega/\partial k_{x}^2$. 

As shown in Fig. \ref{fig:MieResonance}(c), for all PhC structures, the mode profiles of single unit cell resonances are identical to those of BICs occurring in infinite unit cells. The resonance wavelengths of single unit cells are also close to those of BICs. However, there is a trend that a single unit cell resonance with a higher Q-factor has a resonance wavelength closer to the BIC wavelength. Furthermore, if a single unit cell resonance possesses a higher Q-factor, the corresponding PhC structure has a flatter wavelength dispersion, as shown in Fig. \ref{fig:MieResonance}(d). 
This can be explained as follows. Each unit cell works as an individual resonator with a characteristic Q-factor. As unit cells are brought closer to one another, a band of resonance wavelengths is formed by virtue of mutual interactions, while keeping the properties of individual resonances such as field profile and resonance wavelength. This is analogous to the electronic band structure in the tight-binding model of semiconductors. The interaction strength that is inversely proportional to the resonance Q-factor, determines the width of the band. Thus, with the resonance Q-factor higher, the bandwidth is smaller, leading to a flatter band dispersion. 
The enormous Q-factor increase at a specific BIC wavelength with the number of unit cells approaching an infinity, can be explained as the consequence of successive resonance trapping phenomenon \cite{Persson2000}. It can be concluded that the BICs with an ultraflat dispersion are related to both the properties of an individual unit cell as well as the collective effect of many unit cells. 
Since the properties of individual resonances are strongly maintained in an infinite structure, the frequency and Q-factor dispersion curves obtained for infinite structures can be applied to a finite structure as approximations. More rigorous proof will be provided elsewhere.

The high-Q resonance of the individual unit cell is attributed to the destructive interference of two waveguide modes in the single unit cell. 
Each unit cell is seen as a waveguide along the $z$-direction with terminations, as shown in Fig. \ref{fig:MieResonance}(e), which supports a few guided modes WG$_i$. At the terminations, each waveguide mode not only reflects back into itself but also couples to each other as well as the surrounding continuum of the modes as shown schematically in Fig. \ref{fig:MieResonance}(e). Resonances occur in cases that the Fabry-Perot phase condition is satisfied for each mode, when the coupling with other waveguide modes is typically weak \cite{Landreman2016}. However, for a specific thickness and width, the two waveguide modes with the same spatial symmetry, e.g. WG$_1$ and WG$_3$, can couple strongly together at the interface, while their coupling to the continuum of modes destructively interfere. It results in energy building up inside the waveguide, i.e.,  a strong resonance of the single unit cell.

\begin{figure}[t!]
	\centering
	\includegraphics[width=0.42\textwidth]{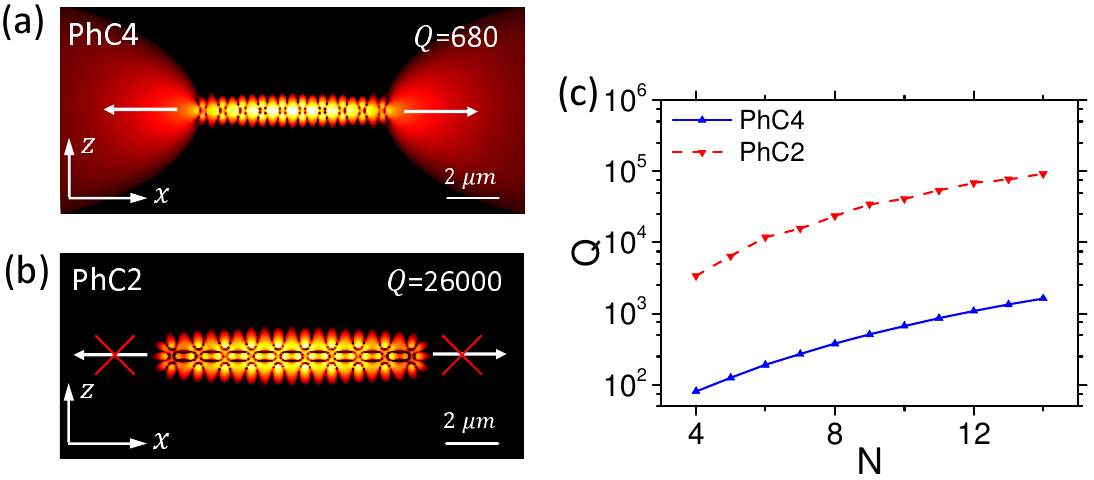}
	\caption{(a),(b) The normalized field profile ($|E_y|$ in dB scale) of the fundamental cavity mode formed by 10 unit cells of PhC4 and PhC2 designs, respectively. (c) The Q-factor of the fundamental cavity mode versus the number of PhC unit cells $N$ for PhC4 (blue solid) and PhC2 (red dashed)
	\label{fig:LateralLoss}}
\end{figure}

Now, we can investigate the impact of an ultraflat frequency dispersion on the in-plane loss of a microcavity formed by a few unit cells of PhC slab, as depicted in Fig. \ref{fig:CavitySchematic}(a). The in-plane loss and corresponding Q-factor, $Q_\parallel$, depend on the group velocities of wavevector components within the $k$-space mode profile (Fourier transform of a coordinate-space mode profile) \cite{Ryu2003, Taghizadeh2015a} and the cavity termination at its boundaries \cite{Xu2005}. For microcavities of a few microns long, this in-plane loss is considerable due to the short length.
The ultraflat dispersion BIC of this paper allows for very small group velocities over all wavevector components, significantly reducing the in-plane loss. 
Figures \ref{fig:LateralLoss}(a) and \ref{fig:LateralLoss}(b) compare the mode profiles of two microcavities based on PhC2 and PhC4.
The PhC2-based microcavity with a smaller dispersion curvature has a much smaller in-plane loss than the PhC4-based one, while both microcavities have negligible out-of-plane loss. This is also shown in Fig. \ref{fig:LateralLoss}(c) which presents the Q-factors of both microcavities as a function of the number of unit cells, $N$. Finally, it is noteworthy to mention that the in-plane Q-factor can be further-enhanced by introducing a heterostructure, which has been widely employed in other form of PhC microcavities \cite{Istrate2006}.

Let us consider the mechanism of out-of-plane loss in the microcavity formed by few unit cells, and a method to minimize it based on $k$-space engineering. In an infinite PhC slab, the $k$-space mode profile of a BIC is a delta function centered at $k_\parallel$=$k_\text{BIC}$, whereas the $k$-space profile of a resonance in a finite structure has a considerable width. The $k$-space components away from $k_\text{BIC}$ couple to the radiation modes above the light line.
Thus, the resultant out-of-plane loss and corresponding Q-factor, $Q_\perp$, depend on the overlap of the $k$-space mode profile of the microcavity with the Q-factor dispersion of the infinite PhC slab. Given a $k$-space mode profile, of which the width is inversely proportional to the microcavity lateral length $L$=$N\Lambda$, the shape of the Q-factor dispersion can be engineered to give a better overlap by controlling the positions of paired non-symmetry-protected BICs, and furthermore by combining them with a symmetry-protected BIC.

\begin{figure}[t!]
	\centering
	\includegraphics[width=0.48\textwidth]{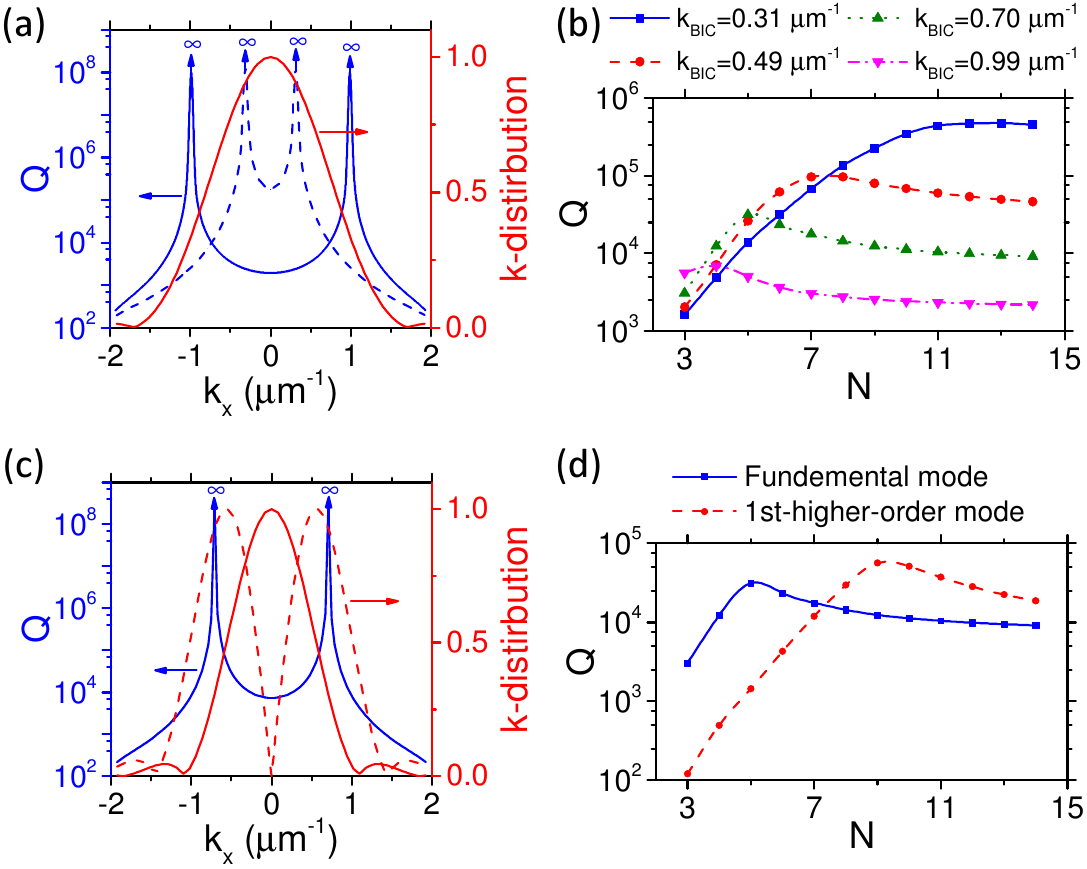}
	\caption{(a) The Q-factor dispersions (blue) of two PhC structures with different $k_\text{BIC}$, and the $k$-space fundamental-mode profile (red) obtained for a microcavity with 6 unit cells of PhC2. (b) The Q-factor of the fundamental mode versus the number of unit cells $N$ for four $k_\text{BIC}$ values. (c) The Q-factor dispersion for a case of $k_\text{BIC}$=0.70~$\mu\text{m}^{-1}$ (blue) and the $k$-space fundamental mode profile (red solid) and 1st-higher-order mode (red dashed) profiles when $N$=9. (d) The Q-factor of the fundamental mode (blue) and 1st-higher-order mode (red) versus $N$. 
	\label{fig:VerticalLoss}}
\end{figure}

The position of paired non-symmetry-protected BICs, $k_\text{BIC}$ can be engineered by changing the PhC parameters, e.g., period, filling ratio, thickness, and refractive index \cite{Zhen2014}. For instance, Fig. \ref{fig:VerticalLoss}(a) illustrates the Q-factor dispersions of two PhC slabs, both based on PhC2  but with slightly different thicknesses, as well as the $k$-space profile of the fundamental mode for a  microcavity with 6 unit cells of PhC2. 
If the BIC positions $\pm k_\text{BIC}$ are too close to $k_x$=0 (blue dashed), most of the $k$-space mode profile overlaps the low Q region of the Q-factor dispersion, leading to a considerable out-of-plane loss. On the other hand, if the BIC positions are far from  $k_x$=0 (blue solid), the Q-factor value becomes low around $k_x$=0 where most $k$-space mode-profile components reside. This results in a poor overlap and consequently a huge out-of-plane loss and small $Q_\perp$. Thus, there is an optimum value of $k_\text{BIC}$ to maximize $Q_\perp$, for a given lateral-size of a microcavity, as shown in Fig. \ref{fig:VerticalLoss}(b). Furthermore, by using cavity modes with different shapes of mode profiles, a better overlap with the Q-factor dispersion can be obtained. This is illustrated in Fig. \ref{fig:VerticalLoss}(c), in which the Q-factor dispersion of infinite PhC and the $k$-space mode profiles of the fundamental (red solid) and 1st-higher-order (red dashed) modes of a cavity with $N$=9 unit cells of that PhC are shown. A larger $Q_\perp$ is expected for the latter due to the better alignment of its $k$-space mode profile with the Q-factor dispersion curve. Figure \ref{fig:VerticalLoss}(d) shows that for $N$$\geq$8, the 1st-higher-order mode has higher Q-factors, as expected. For $N$$<$8, it has more in-plane loss than the fundamental mode, which leads to smaller Q-factors. 


A symmetry-protected BIC can be combined with paired non-symmetry-protected BICs while maintaining the frequency dispersion ultraflat, as shown in Fig. S2.
The two non-symmetry-protected BICs at two off-$\Gamma$-points broaden the high-Q region size, while one symmetry-protected BIC at the $\Gamma$-point elevates the Q-factor value in the middle of the high-Q region. It needs to be noted in Fig. S2(b) (see Supplement 1) that the odd symmetry of the fundamental mode makes its $k$-space mode profile having a node at $k_x$=0. This leads to an excellent matching of the $k$-space profile with the Q-factor dispersion. For example, the high-Q cavity of 3 unit cells shown in Fig. \ref{fig:CavitySchematic}(b) is based on this approach. Another example with 2 unit cells is also illustrated in Fig. S2(b) (see Supplement 1).


The proposed concepts can be employed for designing the other in-plane direction ($y$) as well to make a 3D compact microcavity, as depicted in Fig. \ref{fig:Field3DCase}(a). 
As shown in Fig. \ref{fig:Field3DCase}(b), four non-symmetry-protected and one symmetry-protected BICs form a large area of high Q region in the vicinity of $\Gamma$-point, and give rise to a small out-of-plane loss. Furthermore, the ultraflat dispersion surface results in a small in-plane loss. All these lead to a Q-factor of 1.8$\times10^4$ for a 4 unit cells 3D cavity, of which footprint is less than 20 $\mu\text{m}^2$. The fields are well confined in all three directions, as shown in Figs. \ref{fig:Field3DCase}(c) to \ref{fig:Field3DCase}(e). Another 3D microcavity example based on only non-symmetry-protected BICs is also presented in Fig. S3 (c.f. Supplement 1). In addition, the results in Figs. S4(a) and S4(b) (c.f. Supplement 1) show that high Q-factor microcavities based on BIC can be formed for TM polarization as well as elliptical bars. Though not optimized, these results show that the proposed concepts can be generalized. 

The ultracompact footprint, high field intensity in the air or dielectric, and access to free-space modes of the proposed quasi-BICs may open novel application chances for lasers, sensors, and switches. For instance, the high field intensity in the air for the TM polarization microcavity [c.f. Fig. \ref{fig:Field3DCase}(f)] is highly desirable for sensing of gas molecules or bacteria \cite{Li2011}. 
The strong field at the dielectric surface is ideal for surface-enhanced Raman spectroscopy \cite{Chang2012}, while the strong field enhancement inside the dielectric can be used for strong non-linear effects, e.g., four-wave mixing \cite{Pu2016}. Ultrasmall lasers can also be realized by integrating a gain material inside the microcavity \cite{Ek2014} or hybrid integration of dye-doped organic materials surrounding the cavity \cite{Korn2016}. 



We emphasize here that the very slow group velocities of the BICs of this study is fundamentally different from the slow light in conventional photonic crystals \cite{Krauss2007, Baba2008}. The former originates from the single cell resonance and thus can be effective even for two or three unit cells, whereas the latter results mainly from the periodic interactions among unit cells, being observed for many unit cells. Furthermore, it is also noted that the effect of the BIC phenomenon defined for infinite structure can be kept considerably strong for finite structure with few unit cells by matching the reciprocal-space properties. The resultant possibility of realizing quasi-BICs in a very compact platform may ease the experimental studies of BICs, as well as novel functionalities for important applications such as high-Q microcavities.

\medskip
\noindent {\Large \textbf{Funding}.} Innovation Fund Denmark through the HOT project (Grant No. 5106-00013B).

\medskip
\noindent {\Large \textbf{Acknowledgments}.} The authors thank Prof. Andrei Lavrinenko and Prof. J. M{{\o}rk for helpful discussions.
	
\medskip
\noindent See Supplement 1 for supporting content.

\bibliography{FrozenBIC}

\ifthenelse{\boolean{shortarticle}}{%
	\clearpage
	\bibliographyfullrefs{FrozenBIC}
}{}


\end{document}